\documentclass[article]{jss_edited}

\usepackage{thumbpdf,lmodern}
\usepackage{mixedgcImp}

\usepackage{framed}

\newcommand{\gcpackage}{\pkg{gcimpute} }
\newcommand{\gcpackageEND}{\pkg{gcimpute}}

\author{Yuxuan Zhao\\Cornell University
   \And Madeleine Udell\\Cornell University}
\Plainauthor{Yuxuan Zhao, Madeleine Udell}

\title{gcimpute: A Package for Missing Data Imputation}
\Plaintitle{gcimpute: A Package for Missing Data Imputation}
\Shorttitle{gcimpute in \proglang{Python}}

\Abstract{
  This article introduces the \proglang{Python} package \gcpackage for missing data imputation.
  \gcpackage can impute missing data with many different variable types, 
  including continuous, binary, ordinal, count, and truncated values, 
  by modeling data as samples from a Gaussian copula model.
  This semiparametric model learns the marginal distribution of each variable to match the empirical distribution, yet describes the interactions between variables with a joint
  Gaussian that enables fast inference, imputation with confidence intervals, and multiple imputation.
  The package also provides specialized extensions to handle large datasets (with complexity linear in the number of observations) and streaming datasets (with online imputation).
  This article describes the underlying methodology and demonstrates how to use the software package.
}

\Plainkeywords{missing data, single imputation, multiple imputation, Gaussian copula, mixed data, imputation uncertainty, Python}

\Address{
  Yuxuan Zhao\\
  Department of Statistics and Data Science\\
  Cornell University\\
  Ithaca, NY 14850, United States of America \\
  E-mail: \email{yz2295@cornell.edu}\\
  URL: \url{https://sites.coecis.cornell.edu/yuxuanzhao/}

\vspace{0.75cm}

  Madeleine Udell\\
  Operation Research and Information Engineering\\
  Cornell University\\
  Ithaca, NY 14850, United States of America \\
  E-mail: \email{udell@cornell.edu}\\
  URL: \url{https://people.orie.cornell.edu/mru8/}
}

\begin{document}



\section{Introduction} \label{sec:intro}
Missing data is ubiquitous in modern datasets, yet most machine learning algorithms and statistical models require complete data.
Thus missing data imputation forms the first critical step of many data analysis pipelines.
The difficulty is greatest for mixed datasets, 
including continuous, binary, ordinal, count, nominal and truncated variables.
Mixed datasets may appear either as a single dataset recording different types of attributes or an integrated dataset from multiple sources.
For example, social survey datasets are generally mixed since they often contain age (continuous), demographic group variables (nominal), and Likert scales (ordinal) measuring how strongly a respondent agrees with certain stated opinions, such as the five category scale: strongly disagree, disagree, neither agree or disagree agree, strongly agree.
The Cancer Genome Atlas Project is an example of integrated mixed dataset: 
it contains gene expression (continuous), mutation (binary) and microRNA (count) data.
Imputation may be challenging even for datasets with only continuous variables 
if variables have very different scales and variability.

The Gaussian copula model nicely addresses the challenges of modeling mixed data 
by separating the multivariate interaction of the variables from their marginal distributions \citep{liu2009nonparanormal, hoff2007extending, fan2017high}.
Specifically,
this model posits that each data vector is generated by first drawing a latent Gaussian vector and then transforming it to match the observed marginal distribution of each variable. In this way, ordinals result from thresholding continuous latent variables.
A copula correlation matrix fully specifies the multivariate interaction 
and is invariant to strictly monotonic marginal transformations of the variables.

\citet{zhao2020missing} proposes to impute missing data by learning a Gaussian copula model from incomplete observation and shows empirically the resulting imputation achieves state-of-the-art performance.
Following this line of work,
\citet{zhao2020matrix} develops a low rank Gaussian copula that scales well to large datasets,
and \citet{zhao2020online} extends the model to online imputation of a streaming dataset using online model updates.
This article introduces an additional methodological advance by extending the Gaussian copula model to support truncated variables.
Truncated variables are continuous variables that are truncated to an interval (which may be half-open) (see \cref{sec:model} and \cref{tab:marginal} for precise definition).
One example is the zero-inflated variable: a non-negative variable with excess zeros,
which often appears when a continuous variable is measured by a machine that cannot distinguish small values from zero.

Reliable decision-making with missing data requires a method to assess the uncertainty 
introduced by imputation.
Typically, imputation software quantifies uncertainty either by providing explicit 
confidence intervals for imputation, or providing multiple imputations \citep{rubin1996multiple}.
Multiple imputations allow the end user to incorporate imputation uncertainty
into subsequent analysis, for example, by conducting the desired analysis on each imputed dataset and combining the results.
\citet{zhao2020matrix} derives analytical imputation confidence intervals when all variables are continuous. 
In this article, we further develop a multiple imputation method for Gaussian copula imputation.
Furthermore, 
we provide  confidence intervals based on multiple imputation that are valid 
for mixed data.

The package \gcpackage implements the methodology presented in \citet{zhao2020missing}, \citet{zhao2020matrix}, \citet{zhao2020online} and the new advances mentioned above:
it supports imputation for continuous, binary, ordinal, count, and truncated data,
confidence intervals,
multiple imputation,
large-scale imputation using the low rank Gaussian copula model, and online imputation.
Nominal variables cannot be directly modeled by a Gaussian copula model, but \gcpackage also accepts nominal variables by one-hot encoding them into binary variables.
We present the technical background in \cref{sec:models} and demonstrate how to use  \gcpackage through examples drawn from real datasets in \cref{sec:illustrations}.

\subsection{Software for missing data imputation}
Many software implementations are available for missing data imputation. 
The options are most plentiful in \proglang{R}\footnote{See \url{https://cran.r-project.org/web/views/MissingData.html}, \url{https://rmisstastic.netlify.app/rpkg/}.}.
In contrast, most advanced \proglang{Python} imputation packages re-implement earlier \proglang{R} packages.
Hence we restrict our discussion here to \proglang{R} packages.
\gcpackage is available in both languages.


An imputation package will tend to work best on data that matches the 
distributional assumptions used to develop it.
The popular package \pkg{Amelia} \citep{Amelia} makes the strong assumption that the input data is jointly normally distributed,
which cannot be true for mixed data.
\pkg{missMDA} \citep{missMDA} imputes missing data based on principal component analysis,
and handles mixed data by one hot encoding nominal variables.
\pkg{MICE} \citep{mice} and \pkg{missForest} \citep{missForest} iteratively train models to predict each variable from all other variables. They handle mixed data by choosing appropriate learning methods based on each data type.
\pkg{missForest} uses random forest models as base learners, and so yields more accurate imputations than \pkg{MICE}, which uses variants of linear models \citep{stekhoven2012missforest}.
In the computational experience of the authors, 
\gcpackage outperforms \pkg{missForest} on binary, ordinal and continuous mixed data \citep{zhao2020missing}.
When the data includes nominal variables, which are poorly modeled by 
any of the other assumptions (low rank, joint normality, or Gaussian copula), \pkg{missForest} generally works best.

\pkg{Amelia}, \pkg{MICE}, and \pkg{missForest} can work well when the number of variables $p$ is small,
but run too slowly for large $p$.
When the number of samples $n$ is large, methods with weaker structural assumptions like 
\gcpackage and \pkg{missForest} yield better imputations,
as they are able to learn more complex relationships among the variables.
Methods that rely on a low rank assumption scale well to large datasets.
They tend to perform well when 
both $n$ and $p$ are large, 
as these data tables generally look approximately low rank \citep{udell2019big},
but can fail when either $n$ or $p$ is small.
Low rank imputation methods include \pkg{missMDA}, \pkg{softImpute} \citep{softImpute}, \pkg{GLRM} \citep{udell2016generalized} and the low rank model from \gcpackageEND.
Hence \gcpackage provides a compelling imputation method for data of all moderately large sizes.

There are also a few copula based imputation packages in \proglang{R}. \pkg{sbgcop} \citep{sbgcop} uses the same model as \gcpackage but provides a Bayesian implementation using a Markov Chain Monte Carlo (MCMC) algorithm.
\gcpackage uses a frequentist approach to achieve the same level of accuracy as \pkg{sbgcop} much more quickly \citep{zhao2020missing}. 
\pkg{mdgc} \citep{mdgc} amends the algorithm in \citet{zhao2020missing} by 
using a higher quality approximation for certain steps in the computation,
improving model accuracy but significantly increasing the runtime when the number of variables is large ($n>100$).
\pkg{CoImp} \citep{CoImp} uses only complete cases to fit the copula model and is unstable when most instances have missing values.
In contrast, \gcpackage can robustly fit the model even when every instance contains missing values.
Moreover, \gcpackage is the first copula package to fit extremely large datasets (large $p$), by assuming the copula has low rank structure, and the first to fit streaming datasets, using online model estimation. 



\section{Mathematical background} 
\label{sec:models}
\gcpackage fits a Gaussian copula model on a data table with missing entries and uses the fitted model to impute missing entries.
It can return a single imputed data matrix with imputation confidence intervals, 
or multiple imputed data matrices. 
Once a Gaussian copula model is fitted,
it can also be used to impute missing entries 
in new out-of-sample rows.

Let's imagine that we wish to use \gcpackage
on a data table $\mathbf{X}$ with $n$ rows and $p$ columns. 
We refer to each row $\mathbf{x}$ of $\mathbf{X}$ as a sample,
and each column as a variable.
\gcpackage is designed for datasets whose variables admit a total order: 
that is, for any two values of the same variable $x_1$ and $x_2$, either $x_1 > x_2$ or $x_1 \leq x_2$.
Each variable may have a distinct type: for example,
numeric, boolean, ordinal, count, or truncated.
Nominal variables do not have an ordering relationship.
By default, \gcpackage encodes nominal variables as binary variables using a one-hot encoding, 
although other encodings are possible.
\gcpackage learns the distribution of each variable 
in order to better model the data.

\gcpackage offers specialized implementations for large datasets and streaming datasets.
Large datasets with many samples or many variables
can use an efficient implementation that exploits mini-batch training, parallelism, and
low rank structure. 
For streaming datasets,
it can impute missing data immediately upon seeing a new sample 
and update model parameters 
without remembering all historical data.
This method is more efficient and can offer a better fit for non-stationary data.

\subsection{Gaussian copula model}\label{sec:model}
The Gaussian copula \citep{hoff2007extending,liu2009nonparanormal,fan2017high,feng2019high,zhao2020missing} models complex multivariate distributions as tranformations of latent Gaussian vectors.
More specifically, it assumes that the complete data $\bx\in\Rbb^p$ is generated as a monotonic transformation of a latent Gaussian vector $\bz$: 
\[
\bx = (x_1,\ldots,x_p) = (f_1(z_1),\ldots, f_p(z_p)):= \bigf(\bz), \mbox{ for }\bz \sim \mathcal{N}(\bo, \Sigma).
\]
The \emph{marginal} transformations $f_1,\ldots, f_p: \Rbb \rightarrow \Rbb$ match the distribution of the observed variable $\bx$ to the transformed Gaussian $\bigf(\bz)$ and are uniquely identifiable given the cumulative distribution function (CDF) of each variable $x_j$.
This model separates the multivariate interaction from the marginal distribution, as the monotone $\bigf$ establishes the mapping from the latent variables to the observed variables while $\Sigma$ fully specifies the dependence structure.
We write $\bx \sim \gc$ to denote that $\bx$ follows the Gaussian copula model with marginal $\bigf$ and copula correlation $\Sigma$.

\begin{figure}[t!]
\centering
\includegraphics[width=\textwidth]{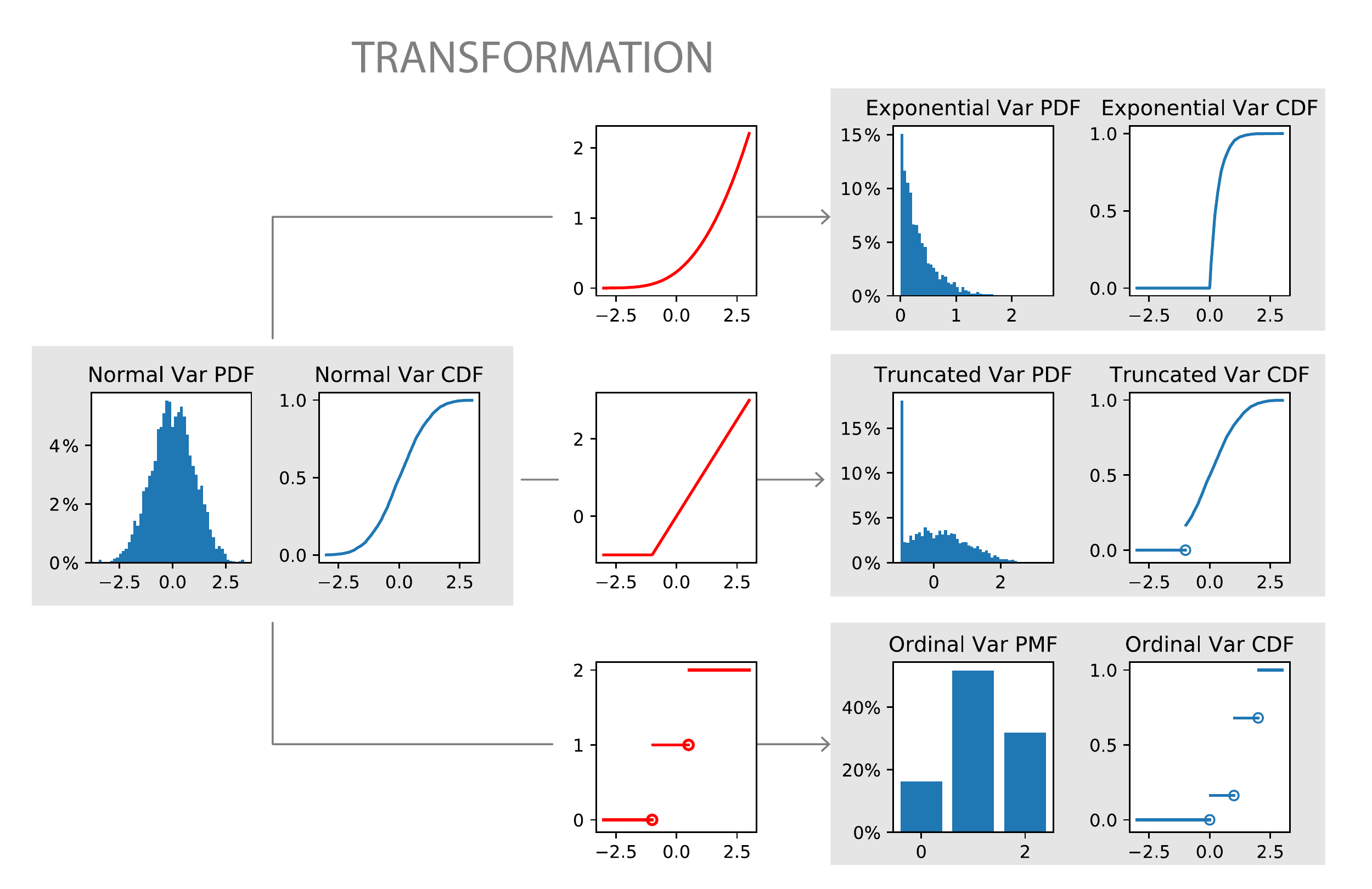}
\caption{\label{fig:marginal_visualization} Three monotoic transformations of a Gaussian variable. 
The third column depicts the transformations that map the data distribution, visualized as both PDF (histogram approximation) and CDF (analytical form), in the left two columns to the data distribution in the right two columns. }
\end{figure}

\begin{figure}[t!]
\centering
\includegraphics[width=\textwidth]{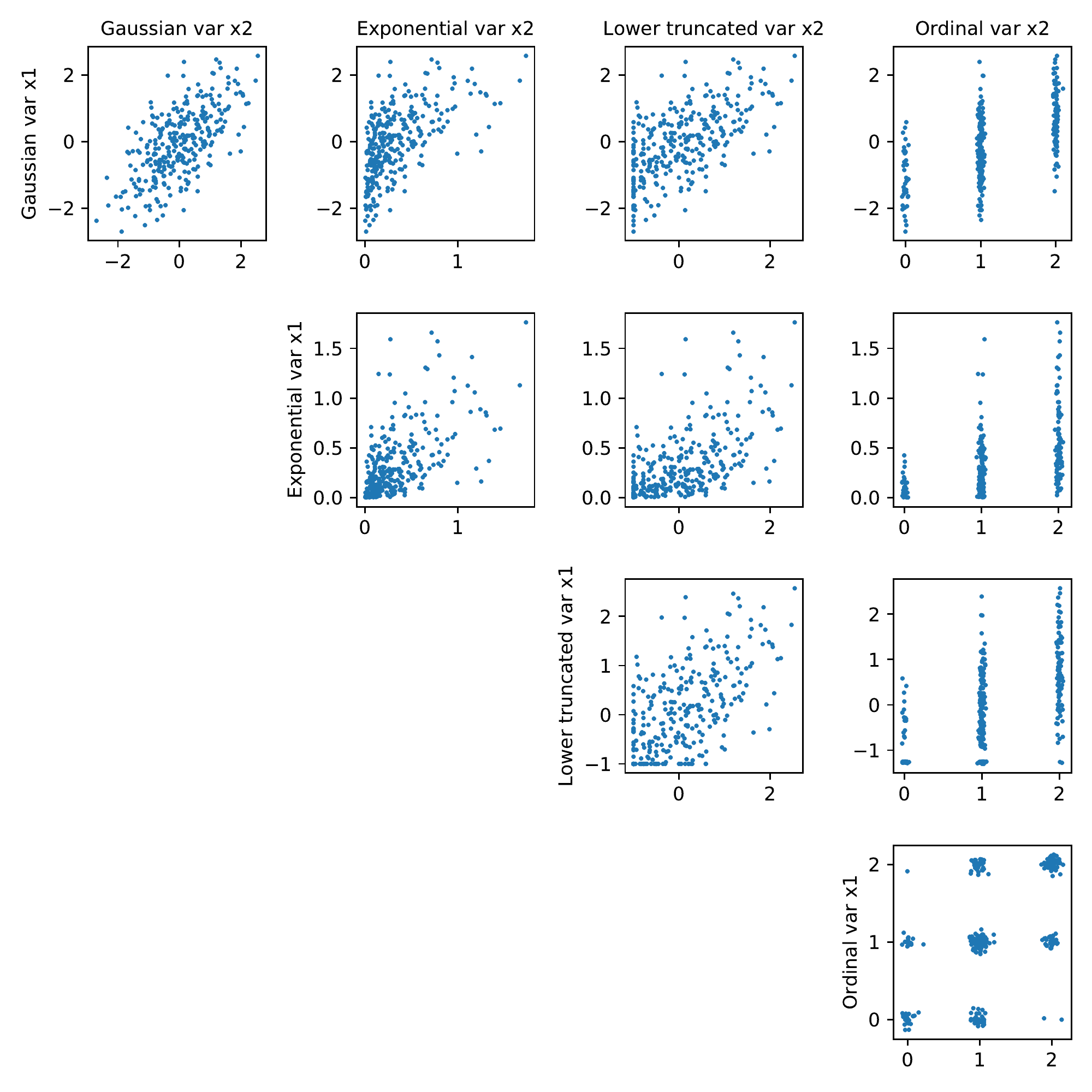}
\caption{\label{fig:dependency_visualization} Scatterplot of samples from several 2D Gaussian copula models with different marginals. 
The data is generated by sampling $(z_1, z_2)$ from a 2D Gaussian distribution with zero mean, unit variance and $.65$ correlation and computing $x_1=f_1(z_1)$ and $x_2=f_2(z_2)$, where $f_1$ and $f_2$ denote the transformations corresponding to the marginals for each model. 
For Gaussian marginals (1st row and 1st column), the transformation is the identity.
For other marginals, the corresponding transformations are plotted as the third column of \cref{fig:marginal_visualization}.}
\end{figure}

\paragraph{Variables and their marginals.} 
When the variable $x_j$ is continuous, $f_j$ is strictly monotonic.
When the variable $x_j$ is ordinal (including binary as a special case), $f_j$ is a monotonic step function \citep{zhao2020missing}.
The copula model also supports one or two sided truncated variables. A one sided truncated variable is a continuous variable truncated either below or above.
A variable $x$ truncated below at $x=\alpha$ has a CDF:
\begin{equation}
    F(x)=\Prm(x=\alpha)\mathds{1}(x\geq \alpha) + (1-\Prm(x=\alpha))\tilde F(x),
    \label{Eq:cdf_truncated}
\end{equation}
where $\tilde F(x)$ is the CDF of a random variable satisfying $\tilde F(\alpha)=0$.
An upper truncated variable and two sided truncated variable are defined similarly. 
The CDF of a truncated variables is a strictly monotonic function with a step either on the left (lower truncated) or the right (upper truncated) or both (two sided truncated).
The expression of $f_j$ as well as their set inverse $f_j^{-1}(x_j) := \{z_j|f_j(z_j)=x_j\}$ are given in \cref{tab:marginal}. 
In short, $f_j$ explains how the data is generated, while $f_j^{-1}$ denotes available information for model inference given the observed data.
\cref{fig:marginal_visualization} depicts 
how a Gaussian variable is transformed into an exponential variable, a lower truncated variable, and an ordinal variable.
\cref{fig:dependency_visualization} depicts the 
dependency structure induced by a Gaussian copula model:
it plots randomly drawn samples from 2D Gaussian copula model with the same marginal distributions from \cref{fig:marginal_visualization}.
It shows that the Gaussian copula model is much more expressive than the multivariate normal distribution.

\begin{table}[t!]
\centering
\begin{tabular}{lll}
\hline
\textbf{Type} \\ \hline 
Continuous & Distribution & $x$ has CDF $F(x)$. \\ 
& $f(z)$& $F^{-1}(\Phi(z))$     \\
&  $f^{-1}(x)$&$\Phi^{-1}(F(x))$
 \\\hline
Ordinal & Distribution & $x$ has probability mass function $\Prm(x=i)=p_i$, for $i=1,...,k.$ \\
& $f(z)$& $\max\left\{i: \sum_{l=0}^{i-1}p_l\leq \Phi(z) < \sum_{l=0}^{i}p_l\right\}$, with $p_0=0$    \\ 
& $f^{-1}(x)$&$\left\{z:\sum_{l=0}^{x-1}p_l\leq \Phi(z)< \sum_{l=0}^xp_l\right\}$, with $p_0=0$  
\\ \hline
Truncated & Distribution & $x$ is truncated into $[\alpha,\beta]$, with $\Prm(x=\alpha)=p_\alpha$, $\Prm(x= \beta)=p_\beta$, \\
& & and CDF $\tilde F(x)$ conditional on $x\in (\alpha,\beta)$, which satisfies  \\
&& $\tilde F(\alpha)=0$ and $\tilde F(\beta)=1$.\\
& $f(z)$&$\begin{cases} 
\alpha, &\Phi(z)\leq p_\alpha \\
\tilde F^{-1}\left(\frac{\Phi(z)-p_\alpha}{1-p_\alpha-p_\beta}\right), &\Phi(z)\in (p_\alpha,1-p_\beta)\\
\beta, & \Phi(z)\geq 1-p_\beta
\end{cases}$\\
& $f^{-1}(x)$&$\begin{cases} 
\{z:\Phi(z)\leq p_\alpha\}, & x=\alpha \\
\Phi^{-1}\left(p_\alpha+(1-p_\alpha-p_\beta)\tilde F(x)\right), &x\in (\alpha,\beta)\\
\{z:\Phi(z)\geq 1-p_\beta\},  & x=\beta
\end{cases}$\\\hline
\end{tabular}
\caption{\label{tab:marginal}For any random variable $x$ admitting a total order, there exists a unique monotonic transformation $f$ such that $f(z)=x$ for a random standard Gaussian $z$.
For each data type of $x$, this table includes its distribution specification, the marginal $f$, and the set inverse $f^{-1}(x)=\{z:f(z)=x\}$ of the marginal. Three different types of truncated variables are summarized together: 
(1) $\alpha=-\infty$ and $p_\alpha=0$ corresponds to lower truncated $x$; 
(2) $\beta=\infty$ and $p_\beta=0$ corresponds to upper truncated $x$; 
(3) finite $\alpha, \beta$ and positive $p_\alpha, p_\beta$ corresponds to two sided truncated $x$.
$\Phi(\cdot)$ denotes the CDF of a standard normal variable.
}
\end{table}

By default, \gcpackage categorizes a count variable to one of the above variable types base on its distribution (see \cref{sec:basic_usage} for the rule).  
\gcpackage also provides a Poisson distribution modeling for count variables. 
The difference embodies in how to estimate $f_j$ and $f_j^{-1}$.
We defer its discussion to \cref{sec:alg} after introducing marginal transformation estimation.

\subsection{Missing data imputation} \label{sec:imputation}
\gcpackage uses the observed entries, along with the estimated dependence structure of the variables, 
to impute the missing entries.
In this section, let us suppose we have estimates of the model parameters $\bigf$ and $\Sigma$
and see how to impute the missing entries.
We discuss how to estimate the model parameters in the next section. 

Every sample is independent conditional on $\bigf$ and $\Sigma$, so we may independently consider how to impute missing data in each sample.
For a sample $\bx \sim \gc$, 
let us denote the observed variables as $\indexO$ and the missing variables as $\indexM$,
so $\xobs$ is a vector of length $|\indexO|$ that collects the observed entries.
Since $\bx\sim \gc$, there exists $\bz\sim \mathcal{N}(\bo, \Sigma)$ such that $\bx = \bigf(\bz)$.
Our first task is to learn the distribution of the missing entries in the latent space. 
Denote $\Sigma_{I,J}$ as the submatrix of $\Sigma$ with rows in $I$ and columns in $J$.
Now, $\bz$ is multivariate normal, so
\[
\zmis|\zobs \sim \mathcal{N}(\SigmaMO\SigmaOOinv\zobs, \SigmaMM-\SigmaMO\SigmaOOinv\SigmaOM).
\]
We can map this distribution to a distribution on $\xmis$ using the marginal transformation $\bigf$. 

\paragraph{Observations and their latent consequences.} To estimate the distribution of $\zmis$, 
we must model distribution of $\zobs$ using the observed values $\xobs$.
For an observed continuous variable value $x_j$, the corresponding $z_j$ takes value $f_j^{-1}(x_j)$ with probability 1.
For an observed ordinal variable value $x_j$, $f_j^{-1}(x_j)$ is an interval, since $f_j$ is a monotonic step function.
Hence the distribution of $z_j$ condition on $x_j$ is a truncated normal in the interval $f_j^{-1}(x_j)$.
For an observed truncated variable $x_j$, 
$z_j$ takes value $f_j^{-1}(x_j)$ with probability 1 if $x_j$ is not the truncated value, otherwise is a truncated normal in the interval $f_j^{-1}(x_j)$.

If only a single imputation is needed, \gcpackage will first compute the conditional mean of $\zmis$ given $\xobs$, 
and then return the imputation by applying the transformation $\bigf$.
If multiple imputations are needed, \gcpackage will instead sample from the conditional distribution of $\zmis$ given $\xobs$, and then again transform the values through $\bigf$.

\gcpackage can return confidence intervals for any single imputation.
If all observed variables $\xobs$ are continuous, 
$\zobs$ has all probability mass at a single point and thus $\zmis$ has a multivariate normal distribution.
In this scenario, \gcpackage first computes the normal confidence interval and then transform it thorough $\bigf$ to produce a confidence interval for the imputation.
In other cases, \gcpackage computes an approximate confidence interval by assuming that $\zobs$ has all probability mass at its conditional mean given $\xobs$ and then compute the normal confidence interval of $\zmis$ as it does for all continuous variables.
The approximated confidence intervals are still reasonably well calibrated if there are not too many ordinal variables.
Otherwise, \gcpackage provides a safer approach to build confidence intervals by performing multiple imputation and taking a confidence interval on the empirical percentiles of imputed values.

\subsection{Algorithm}\label{sec:alg}
Inference for the Gaussian copula model  estimates the marginal transformations $f_1,\ldots,f_p$, as well as their inverses, and the copula correlation matrix $\Sigma$.
The estimate of the marginal distribution and its inverse relies on the empirical distribution of each observed variable. The marginals may be consistently estimated under a missing completely at random (MCAR) mechanism.
Otherwise, these estimates are generally biased: for example, if larger values are missing with higher probability, the empirical distribution is not a consistent estimate of the true distribution.

Estimating the copula correlation $\Sigma$ is a maximum likelihood estimate (MLE) problem. Estimates for the correlation are consistent under the missing at random (MAR) mechanism, provided the marginals are known or consistently estimated \citep{little2019statistical}.

\paragraph{Marginal transformation estimation.}
As shown in \cref{tab:marginal}, both $f_j$ and $f_j^{-1}$ only depends on the distribution of the observed variable $x_j$. 
Thus to estimate the transformation,
we must estimate the distribution of $x_j$, for example, by estimating the CDF and quantile function (for continuous and truncated) or the probability of discrete values with positive probability mass (for ordinal and truncated).
\gcpackage uses the empirical CDF, quantile function, or discrete probability as estimates.

All imputed values are obtained through estimated $f_j$,
and thus the empirical quantile estimate of $F_j^{-1}$.
For continuous variables, a linear interpolated $f_j$ is used so that the imputation is a weighted average of the observed values.
For ordinal variables, the imputation is a most likely observed ordinal level.

Suppose you know a parametric form of $f_j$ for the observed data. Can you use this information? Should you use this information?
1) Yes, you can use this information. We include a capacity to do this in our package.
2) No, you probably shouldn't. We have never seen a dataset when using the parametric form helps. For example, for Poisson (count) data with a small mean, most likely values are observed, so treating the data as ordinal works well.
For Poisson data with a large mean, the empirical distribution does miss certain values, so certain values will \emph{never} appear as imputations. Yet we find that fitting a parametric form instead barely outperforms! 
We believe that the dangers of model misspecification generally outweigh the advantage of a correctly specific parametric model.
Parametric and nonparametric models differ most in their predictions of tail events. Alas, these predictions are never very reliable: it is difficult to correctly extrapolate the tail of a distribution from the bulk.

\paragraph{Copula correlation estimation.}
\gcpackage uses an expectation maximization (EM) algorithm to estimate the copula correlation matrix $\Sigma$.
Suppose $\bx^1,\ldots,\bx^n$ are $n$ i.i.d. samples from a Gaussian copula model, with observed parts  $\{\xio\}_{i=1,\ldots,n}.$
Denote their corresponding latent Gaussian variables as $\bz^1,\ldots,\bz^n$.
At each E-step, the EM method computes the expected covariance matrix of the latent variables $\bz^i$ given the observed entries $\xio$, i.e. $\frac{1}{n}\sum_{i=1}^n\Erm[\bz^i(\bz^i)^\top|\xio]$ and $\frac{1}{n}\sum_{i=1}^n\Erm[\bz^i|\xio]$ . 
The M-step finds the MLE for the correlation matrix of $\bz^1,\ldots,\bz^n$:
it updates the model parameter $\Sigma$ as the correlation matrix associated with the expected covariance matrix computed in the E-step.
Each EM step has computational complexity $O(np^3)$
(for dense data).

\subsection{Acceleration for large datasets}\label{sec:acceleration_method}
\gcpackage runs quickly on large datasets by exploiting parallelism, mini-batch training and low rank structure to speed up inference.
Our EM algorithm parallelizes easily: the most expensive computation, the E-step, 
is computed as a sum over samples and thus can be easily distributed over multiple cores.

When the number of samples $n$ is large, 
users can invoke mini-batch training to accelerate inference \citep{zhao2020online},
since a small batch of samples already gives an accurate estimate of the full covariance.
This method shuffles the samples, divides them into mini-batches, and uses an online learning algorithm.
Concretely, for $t$-th mini-batch, 
\gcpackage computes the copula correlation estimate, $\hat\Sigma$, using only this batch and then updates the model estimate as 
\begin{equation}
\Sigma^t = (1-\eta_{t})\Sigma^{t-1} + \eta_t\hat \Sigma, 
    \label{Eq:model_update_online}
\end{equation}
where $\Sigma^t$ denotes the correlation estimate and $\eta_t\in (0,1)$ denotes the step size at iteration $t$. To guarantee convergence, the step size $\{\eta_t\}$ must be monotonically decreasing and satisfy $\sum_{t=0}^\infty\eta_t^2<\sum_{t=0}^\infty\eta_t=\infty$.
This online EM algorithm converges much faster as the model is updated more frequently.
\citet{zhao2020online} reports the mini-batch algorithm can reduce train time by up to $85\%$.

When the number of variables $p$ is large, 
users can invoke a low-rank assumption on the covariance
to speed up training.
This low rank Gaussian copula (LRGC) \citep{zhao2020matrix} assumes a factor model for the latent Gaussian variables: 
\begin{equation}
    \bz = W\bt +\beps, \mbox{ where } W\in \Rbb^{p\times k}, \bt \sim \mathcal{N}(0, \Irm_k), \beps \sim \mathcal{N}(0, \sigma^2\Irm_p) \mbox{ with }\sigma^2>0,
    \label{Eq:LRGC}
\end{equation}
for some  rank $k\ll p$. 
The $k$-dimensional $\bt$ is the data generating factors and $\beps$ denotes random noise.
Consequently, the copula correlation matrix has a low rank plus diagonal structure: 
$\Sigma = WW^\top + \sigma^2\Irm_p$.
This factorization decreases the number of parameters from $O(p^2)$ to $O(pk)$ and decreases the per-iteration complexity from $O(np^3)$ to $O(npk^2)$ for dense data.
For sparse data, the computation required is linear in the number of observations.
Thus \gcpackage can easily fit datasets with thousands of variables \citep{zhao2020matrix}.

\subsection{Imputation for streaming datasets}\label{sec:online}
\gcpackage provides an online method to handle imputation in the streaming setting: as new samples arrive, 
it imputes the missing data immediately and then updates the  model parameters.
The model update is similar to offline mini-batch training as presented in \cref{Eq:model_update_online}, with $\hat \Sigma$ estimated from the new samples.
Online imputation methods can outperform offline imputation methods for non-stationary data by quickly adapting to a changing distribution,
while offline methods are restricted to a single, static model.

\gcpackage responds to the changing distribution by updating its parameters $\bigf$ and $\Sigma$ after each sample is observed.
The marginal estimate only uses the $m$ most recent data points, 
so the model forgets stale data and the empirical distribution requires constant memory.
The hyperparameter $m$ should be chosen to reflect how quickly the distribution changes.
A longer window works better when the data distribution is mostly stable but has a few abrupt changes. 
On the other hand, if the data distribution changes rapidly, a shorter window is needed. 
The correlation $\Sigma$ is updated according to the online EM update after observing each new mini-batch, using a constant step size $\eta_t \in (0,1)$. A constant step size ensures the model keeps learning from new data and forgets stale data.

Streaming datasets may have high autocorrelation, 
which can improve online imputation.
By default, \gcpackage imputes missing entries by empirical quantiles of the most recent stored observations.
However, it also supports allocating different weights to different stored observations and imputing missing entries by empirical weighted quantiles.
\gcpackage provides an implementation using decaying weights for the $m$ stored observations: $d^t$ with $d\in (0,1]$ for each time lag $t=1,...,m$.
The decay rate $d$ should be tuned for best performance.
This approach interpolates between imputing the last observed value (as $d\to 0$) 
and the standard Gaussian copula imputation (when $d=1$).
The user may also supply their own choice of weights. 

\section{Software usage} \label{sec:illustrations}
The method presented above for imputing missing data with the Gaussian copula has been implemented in two languages: as the \proglang{Python} package \gcpackage and a R package\footnote{https://github.com/udellgroup/gcimputeR}. Both versions employ the same algorithms and offer the same features. This article demonstrates how to use the \proglang{Python} version. The usage for the \proglang{R} package is similar.
Our examples rely on some basic \proglang{Python}  modules for data manipulation and plotting:
\begin{CodeChunk}
\begin{CodeInput}
>>> import numpy as np
>>> import pandas as pd
>>> import time
>>> import matplotlib.pyplot as plt
>>> import seaborn as sns
>>> from tabulate import tabulate
\end{CodeInput}
\end{CodeChunk}

\subsection{Basic usage}\label{sec:basic_usage}
\begin{figure}[t!]
\centering
\includegraphics[width=\textwidth]{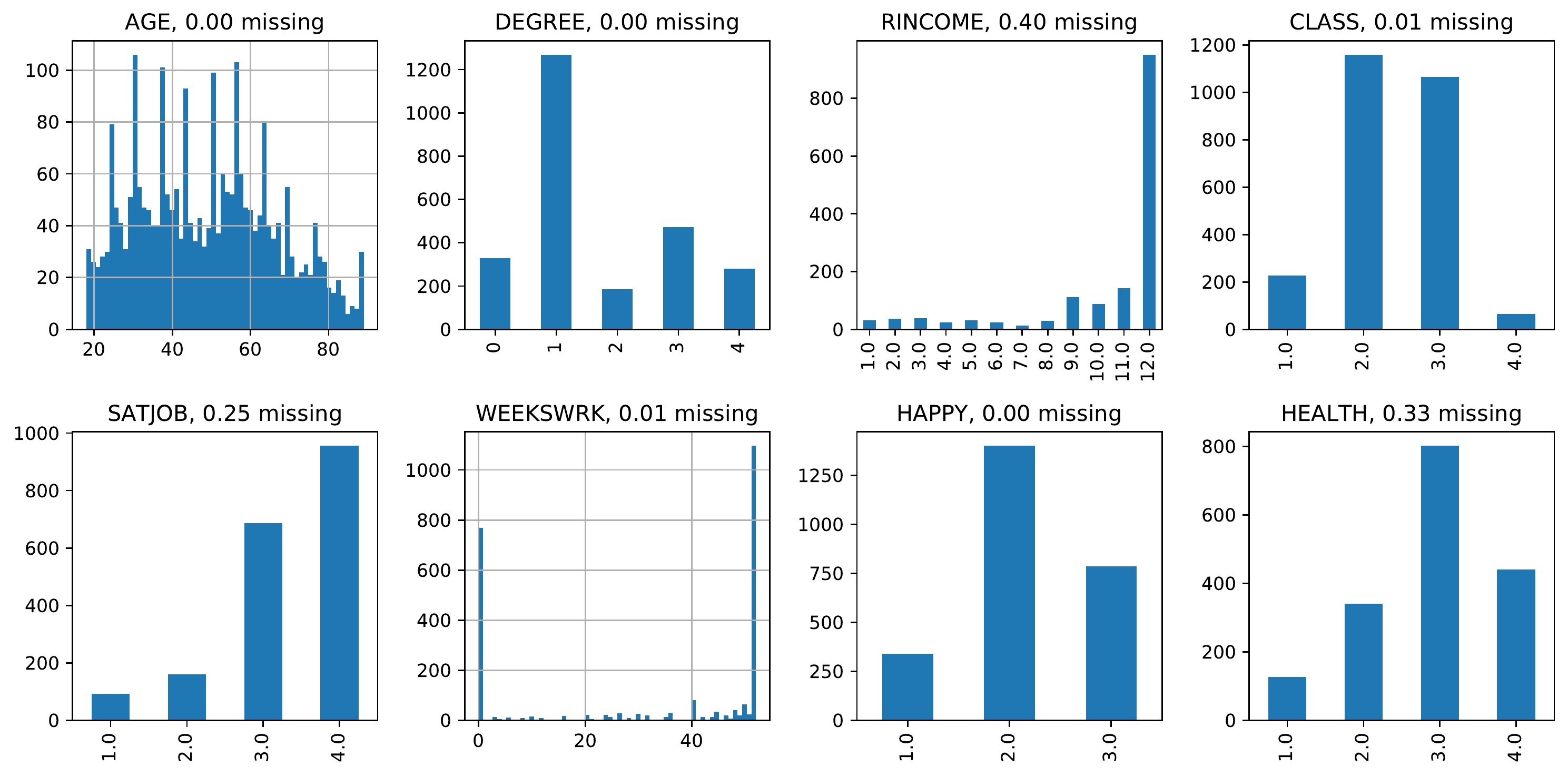}
\caption{\label{fig:GSS_hist} Histogram plots for GSS variables. There are $2538$ samples in total.}
\end{figure}

To demonstrate the basis usage of \gcpackageEND,
we use demographic data from the 2014 General Social Survey (GSS) data: we consider the variables age (\code{AGE}), highest degree (\code{DEGREE}), income (\code{RINCOME}), subjective class identification (\code{CLASS}), satisfaction with the work (\code{SATJOB}), weeks worked last year (\code{WEEKSWRK}), general happiness (\code{HAPPY}), and condition of health (\code{HEALTH}). All variables are ordinal variables encoded as integers, with varying number of ordinal categories. 
The integers could represent numbers, such as $0, 1, \cdots, 52$ for \code{WEEKSWRK}, or ordered categories, such as 1 (``Very happy''), 2 (``Pretty happy''), 3 (``Not too happy'') for the question ``how would you say things are these days'' (\code{HAPPY}).  Many missing entries appear due to answers like ``Don't know'', ``No answer'', ``Not applicable'', etc. Variable histograms are plotted in \cref{fig:GSS_hist} using the following code:
\begin{CodeChunk}
\begin{CodeInput}
>>> from gcimpute.helper_data import load_GSS
>>> data_gss = load_GSS()
>>> fig, axes = plt.subplots(2, 4, figsize=(12,6))
>>> for i,col in enumerate(data_gss):
...    if col in ['AGE', 'WEEKSWRK']:
...        data_gss[col].dropna().hist(ax=axes[i//4, i%4], bins=60)
...    else:
...        to_plot=data_gss[col].dropna().value_counts().sort_index()
...        to_plot.plot(kind='bar', ax=axes[i//4, i%4])
...    _title = f'{col}, {data_gss[col].isna().mean():.2f} missing'
...    axes[i//4, i%4].set_title(_title)
>>> plt.tight_layout()
\end{CodeInput}
\end{CodeChunk}

\begin{figure}[t!]
\centering
\includegraphics[width=0.6\textwidth]{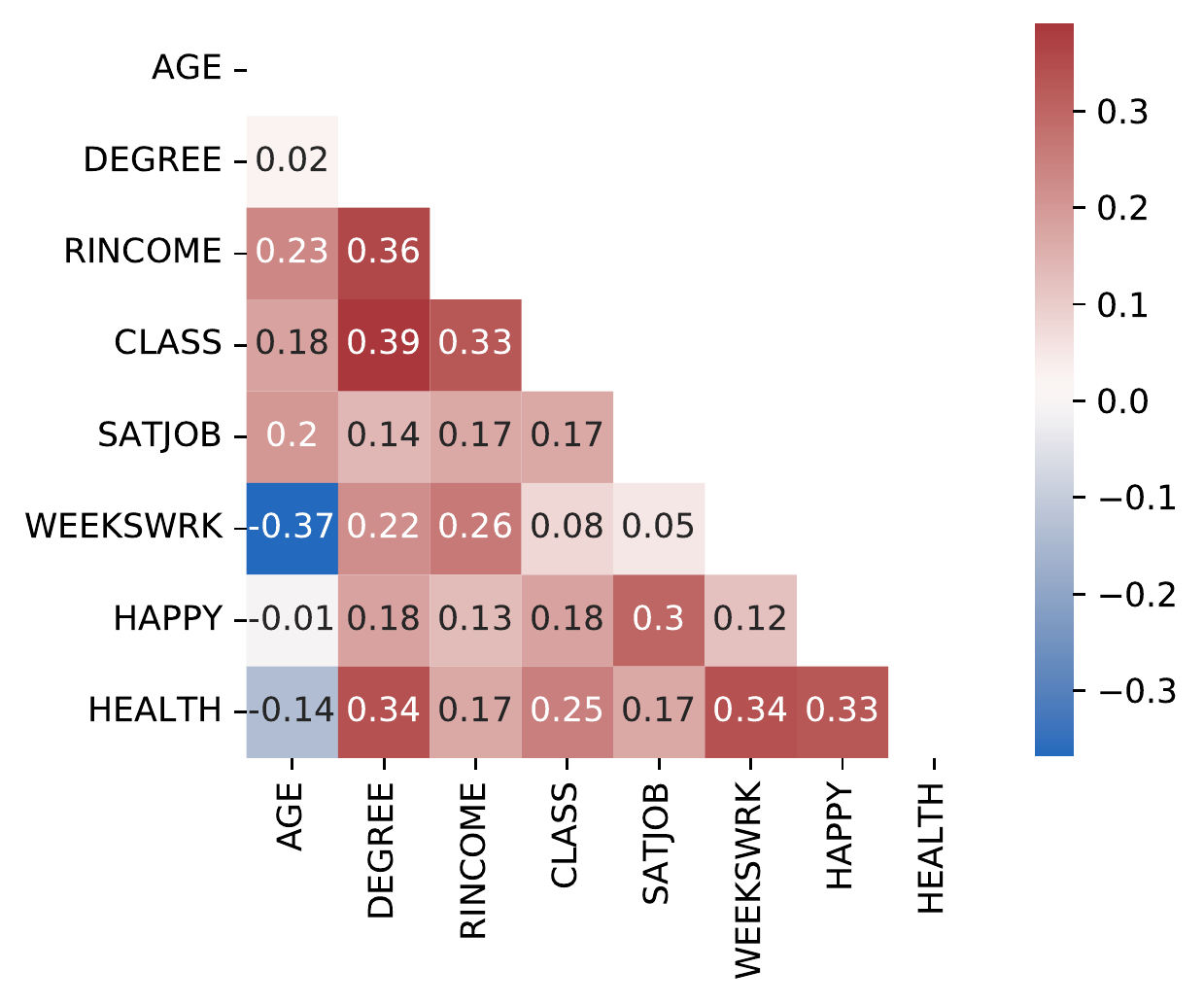}
\caption{\label{fig:GSS_corr} The estimated latent copula correlation among  GSS variables.}
\end{figure}

We mask $10\%$ of the observed entries uniformly at random as a test set to evaluate our imputations. 
\begin{CodeChunk}
\begin{CodeInput}
>>> from gcimpute.helper_mask import mask_MCAR
>>> gss_masked = mask_MCAR(X=data_gss, mask_fraction=.1, seed=101)
\end{CodeInput}
\end{CodeChunk}
The \proglang{Python}  package has an API consistent with the \code{sklearn.impute} module \citep{sklearn_api}. To impute the missing entries in an incomplete dataset, we simply create a model and call \code{fit_transform()}. The default choice uses \code{training_mode=`standard'}, corresponding to the algorithm in  \citet{zhao2020missing}.
\begin{CodeChunk}
\begin{CodeInput}
>>> from gcimpute.gaussian_copula import GaussianCopula
>>> model = GaussianCopula()
>>> Ximp = model.fit_transform(X=gss_masked)
\end{CodeInput}
\end{CodeChunk}
To compare imputation performance across variables with different scales, we use scaled mean absolute error (SMAE) for each variable: the MAE of imputations scaled by the imputation MAE of median imputation.
As shown below, the Gaussian copula imputation improves over median imputation by $10.9\%$ on average.
\begin{CodeChunk}
\begin{CodeInput}
>>> from gcimpute.helper_evaluation import get_smae
>>> smae = get_smae(x_imp=Ximp, x_true=data_gss, x_obs=gss_masked)
>>> print(f'SMAE average over all variables: {smae.mean():.3f}')
\end{CodeInput}
\begin{CodeOutput}
SMAE average over all variables: 0.891
\end{CodeOutput}
\end{CodeChunk}
We can also extract the copula correlation estimates to see which variables are correlated, as in \cref{fig:GSS_corr}.  
Interestingly, 
\code{DEGREE} and \code{CLASS} have the largest positive  correlation $0.39$, 
while \code{WEEKSWRK} and \code{AGE} have the largest negative correlation $-0.37$.
\begin{CodeChunk}
\begin{CodeInput}
>>> copula_corr_est = model.get_params()['copula_corr']
>>> mask = np.zeros_like(copula_corr_est)
>>> mask[np.triu_indices_from(mask)] = True
>>> names = data_gss.columns
>>> sns.heatmap(np.round(copula_corr_est,2), 
                xticklabels=names, yticklabels=names, 
                annot=True, mask=mask, square=True, cmap='vlag')
\end{CodeInput}
\end{CodeChunk}

\subsubsection{Determining the variable types}
The choice of variable type can have a strong effect on inference and imputation.
\gcpackage defines five variable types: \code{`continuous'}, \code{`ordinal'},
\code{`lower_truncated'}, \code{`upper_truncated'} and  \code{`twosided_truncated'}.
\gcpackage provides good default guesses of data types, which we used 
in the previous call.
After fitting the model, we can query the model to ask which variable type was chosen as shown below. 
Only \code{AGE} is treated as continuous; all other variables are treated as ordinal. 
No variable is treated as truncated.
\begin{CodeChunk}
\begin{CodeInput}
>>> for k,v in model.get_vartypes(feature_names=names).items():
>>>     print(f'{k}: {v}')
\end{CodeInput}
\begin{CodeOutput}
continuous: ['AGE']
ordinal: ['DEGREE', 'RINCOME', 'CLASS', 'SATJOB', 'WEEKSWRK', 'HAPPY', 'HEALTH']
lower_truncated: []
upper_truncated: []
twosided_truncated: []
\end{CodeOutput}
\end{CodeChunk}
We can specify the type of each variable 
in \code{model.fit_transform()} directly.
Otherwise, the default setting works well. 
It guesses the variable type based on the frequency of observed unique values.
A variable is treated as continuous if its mode's frequency is less than $0.1$.
A variable is treated as lower/upper/two sided truncated if its minimum's/maximum's/minimum's and maximum's frequency is more than  $0.1$ and the distribution, excluding these values, is continuous by the previous rule. 
All other variables are ordinal.
The default threshold value $0.1$  works well in general, but can be changed using the parameter \code{min_ord_ratio} in the model call \code{GaussianCopula()}.
For example, 
let us look at the frequency of the min, max, and mode for each GSS variable.
\begin{CodeChunk}
\begin{CodeInput}
>>> def key_freq(col):
...    freq = col.value_counts(normalize=True)
...    key_freq = {'mode_freq':freq.max()}
...    _min, _max = col.min(), col.max()
...    key_freq['min_freq'] = freq[_min]
...    key_freq['max_freq'] = freq[_max]
...    freq_middle = freq.drop(index = [_min, _max])
...    key_freq['mode_freq_nominmax'] = freq_middle.max()/freq_middle.sum()
...    return pd.Series(key_freq).round(2)
>>> table = data_gss.apply(lambda x: key_freq(x.dropna())).T
>>> print(tabulate(table, headers='keys', tablefmt='psql'))
\end{CodeInput}
\begin{CodeOutput}
+----------+--------+-------+-------+-----------------+
|          |   mode |   min |   max |   mode_nominmax |
|----------+--------+-------+-------+-----------------|
| AGE      |   0.02 |  0    |  0.01 |            0.02 |
| DEGREE   |   0.5  |  0.13 |  0.11 |            0.66 |
| RINCOME  |   0.62 |  0.02 |  0.62 |            0.26 |
| CLASS    |   0.46 |  0.09 |  0.03 |            0.52 |
| SATJOB   |   0.5  |  0.5  |  0.05 |            0.81 |
| WEEKSWRK |   0.44 |  0.31 |  0.44 |            0.13 |
| HAPPY    |   0.55 |  0.31 |  0.13 |            1    |
| HEALTH   |   0.47 |  0.26 |  0.07 |            0.7  |
+----------+--------+-------+-------+-----------------+
\end{CodeOutput}
\end{CodeChunk}
Only \code{AGE} has mode frequency below $0.1$ and thus is treated as continuous. 
All other variables have strong concentration on a single value,
even after removing the min and max, so these are treated as ordinal.
\code{WEEKSWRK} is an interesting example.
It has 53 levels, yet 75\% of the population works either 0 or 52 weeks per year: 
thus it is not treated as a continuous variable.
Interestingly, if we insist that \code{WEEKWRK} be treated as continuous, the algorithm diverges! We discuss this phenomenon in an online vignette\footnote{https://github.com/udellgroup/gcimpute/blob/master/Examples/Trouble\_shooting.ipynb}.

\subsubsection{Monitoring the algorithm fitting}
\gcpackage considers the model to have converged when the model parameters no longer change rapidly:
It terminates when ${||\Sigma^{t+1}-\Sigma^{t}||_F}/{||\Sigma^{t}||_F}$ falls below the specified \code{tol}, where $\Sigma^t$ is the model parameter estimate at the $t$-th iteration and $||\cdot||_F$ denotes the Frobenius norm. 
In practice, 
the default value \code{tol=0.01} works well and the algorithm  converges in less than 30 iterations in most cases.

Tracking the objective value may also be useful.
The objective value is the marginal likelihood at the observed locations, averaged over all instances. 
When all variables are continuous, 
\gcpackage computes the exact likelihood.
In other cases, 
\gcpackage computes an approximation to the likelihood.
The approximation behaves well in most cases including those with all ordinal variables: 
it monotonically increases during the fitting process and finally converges.

To monitor the parameter update and the objective during fitting, simply set \code{verbose=1} in the model call.
\begin{CodeChunk}
\begin{CodeInput}
>>> model = GaussianCopula(verbose=1)
>>> Ximp = model.fit_transform(X=gss_masked)
\end{CodeInput}
\begin{CodeOutput}
Iteration 1: copula parameter change 0.1168, likelihood -9.6913
Iteration 2: copula parameter change 0.0644, likelihood -9.5869
Iteration 3: copula parameter change 0.0366, likelihood -9.5278
Iteration 4: copula parameter change 0.0220, likelihood -9.4942
Iteration 5: copula parameter change 0.0140, likelihood -9.4744
Iteration 6: copula parameter change 0.0093, likelihood -9.4623
Convergence achieved at iteration 6
\end{CodeOutput}
\end{CodeChunk}
Using a tolerance \code{tol} that is too small can require many more iterations and can cause overfitting.
Hence users may wish to tune \code{tol} for a specific dataset for best performance
using \code{fit_transform_evaluate()}.
This function runs the EM algorithm for specified \code{n_iter} iterations and evaluates the imputed dataset using the provided \code{eval_func} at each iteration.
The function \code{eval_func} should take an imputed dataset as input and output the desired evaluation results.
We can design \code{eval_func}  to evaluate the imputation accuracy or the prediction accuracy of a supervised learning pipeline with the imputed dataset as feature matrix.
For example, to evaluate the mean SMAE of the GSS dataset for up to $15$ iterations, we can run the following code:
\begin{CodeChunk}
\begin{CodeInput}
>>> m = GaussianCopula(verbose=1)
>>> get_err = lambda x: get_smae(x, x_true=data_gss, x_obs=gss_masked).mean()
>>> r = m.fit_transform_evaluate(X=gss_masked, eval_func=get_err, num_iter=15)
>>> plt.plot(list(range(1, 16, 1)), r['evaluation'])
>>> plt.title('Imputation error versus run iterations')
>>> plt.xlabel("Run iterations")
>>> plt.ylabel("SMAE")
\end{CodeInput}
\begin{CodeOutput}
Iteration 1: copula parameter change 0.1168, likelihood -9.6913
Iteration 2: copula parameter change 0.0644, likelihood -9.5869
Iteration 3: copula parameter change 0.0366, likelihood -9.5278
Iteration 4: copula parameter change 0.0220, likelihood -9.4942
Iteration 5: copula parameter change 0.0140, likelihood -9.4744
Iteration 6: copula parameter change 0.0093, likelihood -9.4623
Iteration 7: copula parameter change 0.0063, likelihood -9.4545
Iteration 8: copula parameter change 0.0044, likelihood -9.4494
Iteration 9: copula parameter change 0.0032, likelihood -9.4460
Iteration 10: copula parameter change 0.0023, likelihood -9.4437
Iteration 11: copula parameter change 0.0017, likelihood -9.4421
Iteration 12: copula parameter change 0.0012, likelihood -9.4409
Iteration 13: copula parameter change 0.0009, likelihood -9.4401
Iteration 14: copula parameter change 0.0007, likelihood -9.4396
Iteration 15: copula parameter change 0.0005, likelihood -9.4392
\end{CodeOutput}
\end{CodeChunk}

\begin{figure}[t!]
\centering
\includegraphics[width=0.6\textwidth]{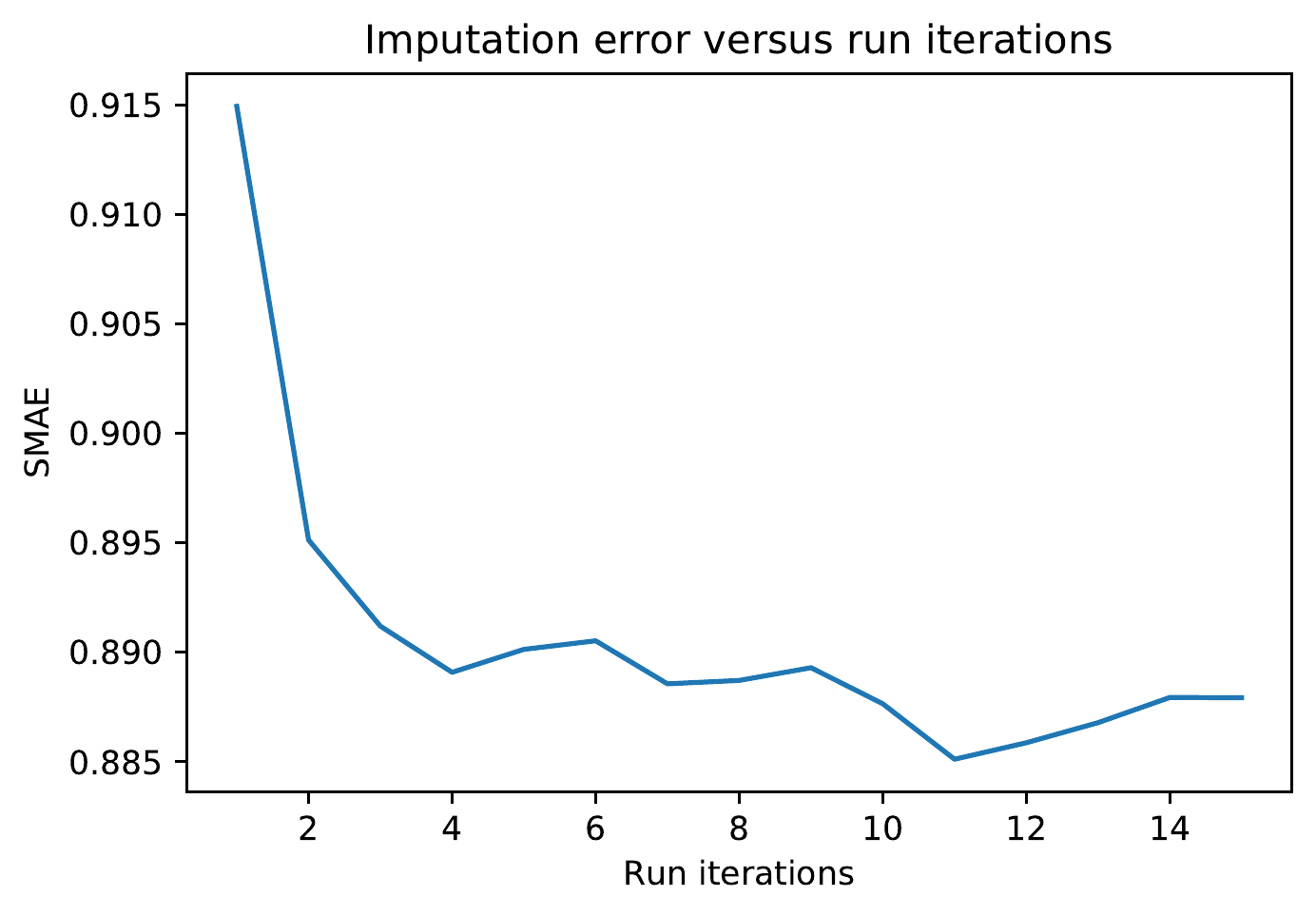}
\caption{\label{fig:GSS_imp_err} The imputation error among GSS variables is plotted w.r.t. the number of iterations run in \gcpackageEND. Satisfactory results emerge after four iterations.}
\end{figure}

Shown in \cref{fig:GSS_imp_err}, the imputation error fluctuates in a small range from $0.885$ to $0.890$ after four iterations.
The default parameter setting stops at iteration $6$.

\subsection{Acceleration for large datasets}
In this section, we will see how to speed up convergence of \gcpackage with 
the acceleration tools described in \cref{sec:acceleration_method}. 
To use parallelism with $m$ cores, we simply set \code{n_jobs=m} in the model call \code{GaussianCopula()}.
To use mini-batching training, we set \code{training_mode} as  \code{`minibatch-offline'} also
in the model call \code{GaussianCopula()}.
The low rank Gaussian copula is invoked using a different model call \code{LowRankGaussianCopula(rank=k)} with desired rank $k$.
Mini-batch training for the low rank Gaussian copula is more challenging and remains for future work,
as the low rank update is nonlinear. 
Nevertheless, for large $n$ and large $p$, the parallel low rank Gaussian copula already converges quite rapidly.

\subsubsection{Accelerating datasets with many samples: mini-batch training}
Mini-batch training requires choosing a decaying step size $\{\eta_t\}$ in \cref{Eq:model_update_online}, a batch size and a maximum number of iterations.
The default setting can be simply invoked by calling \code{GaussianCopula(training_mode=`minibatch-offline')} or explicitly as below:
\begin{verbatim}
model_minibatch = GaussianCopula(training_mode='minibatch-offline', 
                                 stepsize_func = lambda t, c=5:c/(c+t),
                                 batch_size = 100,
                                 num_pass = 2
                                )
\end{verbatim}
The step size sequence $\eta_t$ must satisfy $\eta_t \in (0, 1)$ for all $t$ and $\sum_{t=1}^\infty \eta_t^2<\sum_{t=1}^\infty \eta_t=\infty$.
By default, we recommend using $\eta_t = c/(c+t)$ with $c>0$. 
We find it generally suffices to tune $c$ in the range $(0, 10)$.
The default setting $c=5$ works well in many of our experiments. 

Mini-batch training requires a batch size $s\geq p$ to avoid inverting a singular matrix \citep{zhao2020online}. 
In practice, 
it is easy to select $s\geq p$, since problems with large $p$ should use \code{LowRankGaussianCopula()} instead. 

The maximum number of iterations matters more for mini-batch methods, because the stochastic fluctuation over mini-batches makes it hard to decide convergence based on the parameter update. 
Instead of specifying an exact maximum number of iterations, it may be more convenient to select a desired number of complete passes through the data (epochs), i.e. \code{max_iter}$=\left \lceil{\frac{n}{s}}\right \rceil  \times $\code{num_pass} with $s$ as the mini-batch size. Often using \code{num_pass}$=2$ (the default setting) or $3$ gives satisfying results.

We now run mini-batch training with the defaults on the GSS dataset:
\begin{CodeChunk}
\begin{CodeInput}
>>> t1=time.time()
>>> model_minibatch = GaussianCopula(training_mode='minibatch-offline')
>>> Ximp_batch = model_minibatch.fit_transform(X=gss_masked)
>>> t2=time.time()
>>> print(f'Runtime: {t2-t1:.2f} seconds')
>>> smae_batch = get_smae(x_imp=Ximp_batch, x_true=data_gss, x_obs=gss_masked)
>>> print(f'Imputation error: {smae_batch.mean():.3f}')
\end{CodeInput}
\begin{CodeOutput}
Runtime: 15.27 seconds
Imputation error: 0.886
\end{CodeOutput}
\end{CodeChunk}
Let us also re-run and record the runtime of the standard training mode:
\begin{CodeChunk}
\begin{CodeInput}
>>> t1=time.time()
>>> _ = GaussianCopula().fit_transform(X=gss_masked)
>>> t2=time.time()
>>> print(f'Runtime: {t2-t1:.2f} seconds')
\end{CodeInput}
\begin{CodeOutput}
Runtime: 39.47 seconds
\end{CodeOutput}
\end{CodeChunk}
Mini-batch training not only reduces runtime by $61\%$ but also improves the imputation error (from $0.891$ to $0.886$)! 

\subsubsection{Accelerating datasets with many variables: low rank structure}
The low rank Gaussian copula (LRGC) model accelerates convergence by decreasing the number of 
model parameters. Here we showcase its performance on a subset of the MovieLens1M dataset \citep{harper2015movielens}: 
the $400$ movies with the most ratings and users who rates at least $150$ of these movies in the scale of $\{1,2,3,4,5\}$. 
That yields a dataset consisting of $914$ users and $400$ movies with $53.3\%$ of ratings observed.
We further mask $10\%$ entries for evaluation. 
\begin{CodeChunk}
\begin{CodeInput}
>>> gcimpute.helper_data import load_movielens1m
>>> data_movie = load_movielens1m(num=400, min_obs=150)
>>> movie_masked = mask_MCAR(X=data_movie, mask_fraction=0.1, seed=101)
\end{CodeInput}
\end{CodeChunk}
We run \code{GaussianCopula()} as well as \code{LowRankGaussianCopula(rank=10)}. Here our goal is not to choose the optimal rank, but rather show the runtime comparison between two models. 
\begin{CodeChunk}
\begin{CodeInput}
>>> from gcimpute.low_rank_gaussian_copula import LowRankGaussianCopula
>>> a = time.time()
>>> model_movie_lrgc = LowRankGaussianCopula(rank=10)
>>> m_imp_lrgc = model_movie_lrgc.fit_transform(X=movie_masked)
>>> print(f'LRGC runtime {(time.time()-a)/60:.2f} mins.')
>>> a = time.time()
>>> model_movie_gc = GaussianCopula()
>>> m_imp_gc = model_movie_gc.fit_transform(X=movie_masked)
>>> print(f'GC runtime {(time.time()-a)/60:.2f} mins.')
\end{CodeInput}
\begin{CodeOutput}
LRGC runtime 7.86 mins.
GC runtime 11.66 mins.
\end{CodeOutput}
\end{CodeChunk}
Here we already see that LRGC already reduces the runtime by $34\%$ compared to the standard Gaussian copula,
although the number of variables $p=400$ is not particularly large.
When the number of variables is much larger,
the acceleration is also more significant.
Moreover, LRGC improves the imputation error from $0.616$ to $0.583$, as shown below.
\begin{CodeChunk}
\begin{CodeInput}
>>> from gcimpute.helper_evaluation import get_mae
>>> mae_gc = get_mae(x_imp=m_imp_gc, x_true=data_movie, x_obs=movie_masked)
>>> mae_lrgc = get_mae(x_imp=m_imp_lrgc, x_true=data_movie, x_obs=movie_masked)
>>> print(f'LRGC imputation MAE: {mae_lrgc:.3f}')
>>> print(f'GC imputation MAE: {mae_gc:.3f}')
\end{CodeInput}
\begin{CodeOutput}
LRGC imputation MAE: 0.583
GC imputation MAE: 0.616
\end{CodeOutput}
\end{CodeChunk}
\subsection{Imputation for streaming datasets}
\begin{figure}[t!]
\centering
\includegraphics[width=\textwidth]{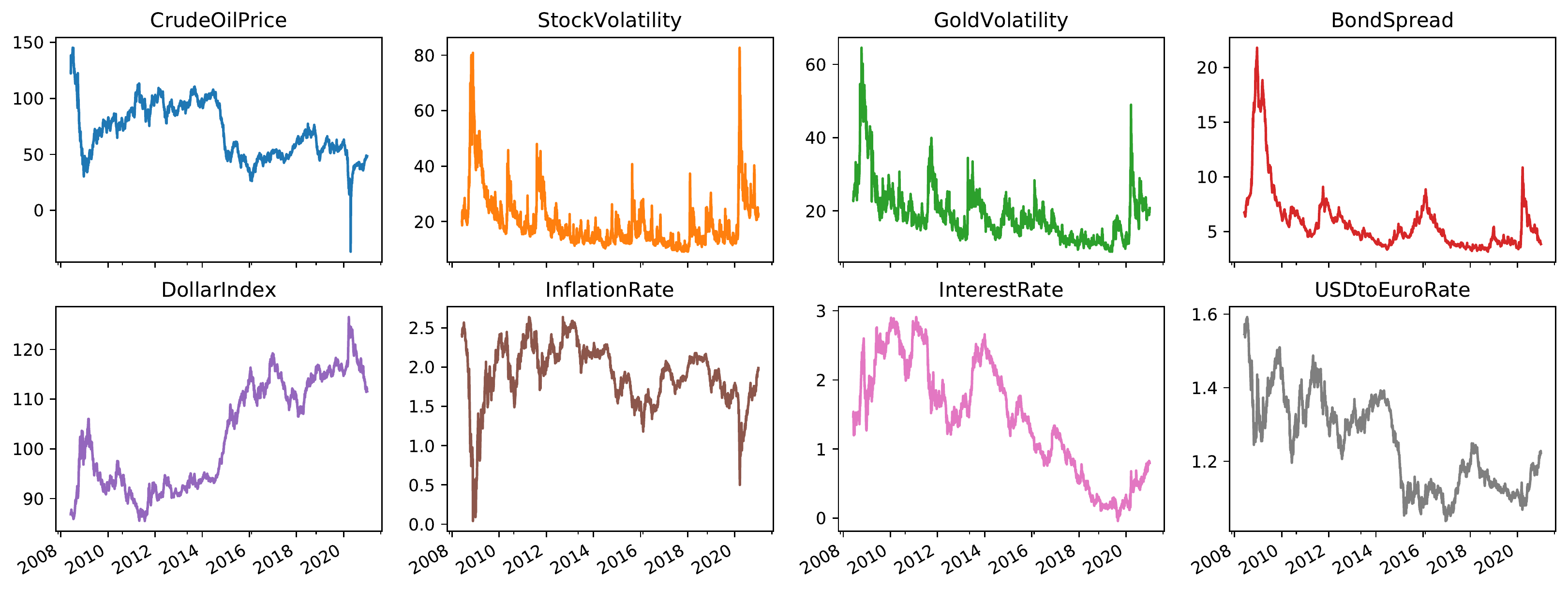}
\caption{\label{fig:fred} Values of eight selected FRED economic variables from 2008-06-03 to 2020-12-31 are plotted.}
\end{figure}

\gcpackageEND's \code{`minibatch-online'} training mode performs streaming imputation:
as new samples arrive, it imputes the missing data immediately and then updates the model parameters.
We showcase its performance on eight daily recorded economic time series variables from federal reserve bank of St. Louis (FRED), consisting of 3109 days from 2008-06-03 to 2020-12-31. The selected eight variables are diverse and among the most popular economic variables in FRED: gold volatility index, stock volatility index, bond spread, dollar index, inflation rate, interest rate, crude oil price, and US dollar to Euro rate,
shown in \cref{fig:fred}.
\begin{CodeChunk}
\begin{CodeInput}
>>> from gcimpute.helper_data import load_FRED
>>> fred_data = load_FRED()
>>> fred_data.plot(subplots = True, layout = (2,4), figsize = (16, 6),
...                legend = False, title = fred_data.columns.to_list()
...               )
\end{CodeInput}
\end{CodeChunk}

Here we consider a scenario in which some variables are observed as soon as they are generated, while others are observed after a lag of one day. 
The goal is to predict the unobserved variables each day. 
We use stock \code{StockVolatility} and \code{CrudeOilPrice} as two unobserved variables. 
Each day,
using a fitted Gaussian copula model, 
we predict their values based on both their historical values (through the marginal) and the six other observed variables at that day (through the copula correlation).
After we make our prediction, the actual values are revealed and used to update the Gaussian copula model.
\gcpackage conveniently supports this task.
Let us first create a Gaussian copula model to impute streaming datasets (\code{training_mode=`minibatch-online'}), shown as below.
\begin{CodeChunk}
\begin{CodeInput}
>>> model = GaussianCopula(training_mode='minibatch-online', 
...                        window_size=10,
...                        const_stepsize=0.1,
...                        batch_size=10,
...                        decay=0.01
...                       )
\end{CodeInput}
\end{CodeChunk}
Three hyperparameters control the learning rate of the model:
\code{window_size} controls the number of recent observations used for marginal estimation;
\code{const_stepsize} controls the size of the copula correlation update;
and \code{batch_size} is the frequency of the copula correlation update.
In contrast, \code{decay} only controls the imputation and does not influence the model update (decay rate $d$ in \cref{sec:online}).
Smaller values of \code{decay} put less weight on old observations, i.e., forget stale data faster.
In economic time series, yesterday's observation often predicts today's value well.
We use a small value \code{decay=.01}, so that the imputation depends most strong on yesterday's observation, but interpolates all values in the window.
These parameters can be tuned for best performance.

Next, to conduct the experiment described above, 
we prepare two data matrices with one row for each temporal observation: 
\code{X} for imputing missing entries 
and \code{X_true} for updating the model.
We use first $25$ rows to initialize the model.
\begin{CodeChunk}
\begin{CodeInput}
>>> fred_masked = fred_data.assign(StockVolatility=np.nan, CrudeOilPrice=np.nan)
>>> Ximp = model.fit_transform(X=fred_masked, X_true=fred_data, n_train=25)
\end{CodeInput}
\end{CodeChunk}
More concretely,
a Gaussian copula model receives the $t$-th row of \code{X}, 
imputes its missing entries, 
and then is asked to update parameters of the model using the $t$-th row of \code{X_true}.
\code{X_true} must agree with \code{X} at all observed entries in \code{X}, but may 
reveal additional entries that are missing in \code{X}.
By default, \code{X_true=None}, indicating no additional entries beyond \code{X} are available.
In this example, two columns of \code{fred_masked} are missing: \code{StockVolatility} and \code{CrudeOilPrice}. All other columns fully observed.
\code{fred_data} has all columns fully observed.

We now evaluate the imputation performance and compare against a simple but powerful alternative, yesterday's observation.
The predicted series of both methods are almost visually indistinguishable from the true values in \cref{fig:fred}, but the Gaussian copula predictions perform better on average,
with lower mean squared error (MSE).
\begin{CodeChunk}
\begin{CodeInput}
>>> n_train = 25
>>> for i, col in enumerate(['CrudeOilPrice', 'StockVolatility']):
...    _true = fred_data[col][n_train:].to_numpy()
...    _err_yes = fred_data[col][n_train-1:-1].to_numpy() - _true
...    _err_GC = Ximp[n_train:,i] - _true
...    print(f'For {col}:')
...    print(f'Gaussian Copula Pred MSE: {np.power(_err_GC,2).mean():.3f}')
...    print(f'Yesterday Value Pred MSE: {np.power(_err_yes,2).mean():.3f}')
\end{CodeInput}
\begin{CodeOutput}
For CrudeOilPrice:
Gaussian Copula Pred MSE: 3.672
Yesterday Value Pred MSE: 4.313
For StockVolatility:
Gaussian Copula Pred MSE: 3.998
Yesterday Value Pred MSE: 4.368
\end{CodeOutput}
\end{CodeChunk}

\subsection{Imputation uncertainty}
So far we have seen several methods to impute missing data.
\gcpackage also provides functionality to quantify the uncertainty of the imputations:
multiple imputation,
confidence interval for a single imputation,
and relative reliability for a single imputation.
We present the first two notions here, since they are widely used.  
The third, relative reliability, aims to rank the imputation quality among all imputed entries \citep{zhao2020matrix}. 
It is well suited for the top-k recommendation task in collaborative filtering.

\subsubsection{Multiple imputation}
Multiple imputation creates several imputed copies of the original dataset, each having potentially different imputed values. The uncertainty due to imputations can be propagated into subsequent analyses by analyzing each imputed dataset. 
Multiple imputation is commonly used in supervised learning when features may have missing entries:
a researcher creates multiple imputed feature datasets, then trains a model with each imputed training feature dataset and predicts with each imputed test feature vector. 
Finally, they pool all predictions into a single prediction, for example, using the mean or majority vote. 
An ensemble model like this often outperforms a single model trained from a single imputation.

We show to use multiple imputation in \gcpackage on a regression task from UCI datasets, the white wine quality dataset \citep{cortez2009modeling}. This dataset has 11 continuous features and a rating target for 4898 samples. The (transposed) header of the dataset is shown below.
\begin{CodeChunk}
\begin{CodeInput}
>>> gcimpute.helper_data import load_whitewine
>>> data_wine = load_whitewine()
>>> print(tabulate(data_wine.head().T, headers='keys', tablefmt='psql'))
\end{CodeInput}
\begin{CodeOutput}
+----------------------+---------+---------+---------+----------+----------+
|                      |       0 |       1 |       2 |        3 |        4 |
|----------------------+---------+---------+---------+----------+----------|
| fixed acidity        |   7     |   6.3   |  8.1    |   7.2    |   7.2    |
| volatile acidity     |   0.27  |   0.3   |  0.28   |   0.23   |   0.23   |
| citric acid          |   0.36  |   0.34  |  0.4    |   0.32   |   0.32   |
| residual sugar       |  20.7   |   1.6   |  6.9    |   8.5    |   8.5    |
| chlorides            |   0.045 |   0.049 |  0.05   |   0.058  |   0.058  |
| free sulfur dioxide  |  45     |  14     | 30      |  47      |  47      |
| total sulfur dioxide | 170     | 132     | 97      | 186      | 186      |
| density              |   1.001 |   0.994 |  0.9951 |   0.9956 |   0.9956 |
| pH                   |   3     |   3.3   |  3.26   |   3.19   |   3.19   |
| sulphates            |   0.45  |   0.49  |  0.44   |   0.4    |   0.4    |
| alcohol              |   8.8   |   9.5   | 10.1    |   9.9    |   9.9    |
| quality              |   6     |   6     |  6      |   6      |   6      |
+----------------------+---------+---------+---------+----------+----------+
\end{CodeOutput}
\end{CodeChunk}
We now randomly mask $30\%$ of entries and fit a Gaussian copula model to the masked dataset.
\begin{CodeChunk}
\begin{CodeInput}
>>> X_wine = data_wine.to_numpy()[:,:-1]
>>> X_wine_masked = mask_MCAR(X_wine, mask_fraction=0.3)
>>> model_wine = GaussianCopula()
>>> X_wine_imputed = model_wine.fit_transform(X=X_wine_masked)
\end{CodeInput}
\end{CodeChunk}
Now we use the first 4000 instances as a training dataset and the remaining 898 instances as test dataset. Since the goal is to show how to use multiple imputation, 
we use simple linear model as the prediction model. 
Now, let us first examine the MSE of the linear model fitted on the complete feature dataset.
\begin{CodeChunk}
\begin{CodeInput}
>>> from sklearn.metrics import mean_squared_error as MSE
>>> from sklearn.linear_model import LinearRegression as LR
>>> X_train, X_test = X_wine[:4000], X_wine[4000:]
>>> y_train, y_test = data_wine['quality'][:4000], data_wine['quality'][4000:]
>>> y_pred = LR().fit(X=X_train, y=y_train).predict(X=X_test)
>>> np.round(MSE(y_test, y_pred),4)
\end{CodeInput}
\begin{CodeOutput}
0.5121
\end{CodeOutput}
\end{CodeChunk}
Now let us examine the MSE of the linear model fitted on the single imputed dataset.
\begin{CodeChunk}
\begin{CodeInput}
>>> X_train_imp, X_test_imp = X_wine_imputed[:4000], X_wine_imputed[4000:]
>>> y_pred_imp = LR().fit(X=X_train_imp, y=y_train).predict(X=X_test_imp)
>>> np.round(MSE(y_test, y_pred_imp),4)
\end{CodeInput}
\begin{CodeOutput}
 0.5295
\end{CodeOutput}
\end{CodeChunk}
Not surprisingly, replacing $30\%$ feature values with the corresponding imputation does hurt the prediction accuracy. Now let us draw $5$ imputed datasets, train a linear model and get prediction for each imputed dataset, and derive the final prediction as the average across 5 different prediction. As shown below, the mean-pooled prediction improves the results from single imputation and performs very close to the results using the complete dataset.
\begin{CodeChunk}
\begin{CodeInput}
>>> X_wine_imputed_mul = model_wine.sample_imputation(X=X_wine_masked, num=5)
>>> y_pred_mul_imputed = []
>>> for i in range(5):
...    X_imputed = X_wine_imputed_mul[...,i]
...    _X_train_imp, _X_test_imp = X_imputed[:4000], X_imputed[4000:]
...    _y_pred = LR().fit(X=_X_train_imp, y=y_train).predict(X=_X_test_imp)
...    y_pred_mul_imputed.append(y_pred_imputed)
>>> y_pred_mul_imputed = np.array(y_pred_mul_imputed).mean(axis=0)
>>> np.round(MSE(y_test, y_pred_mul_imputed), 4)
\end{CodeInput}
\begin{CodeOutput}
 0.5152
\end{CodeOutput}
\end{CodeChunk}
\subsubsection{Imputation confidence intervals}
Confidence intervals (CI) are another important measure of uncertainty.
\gcpackage can return a CI for each imputed value: 
for example, a 95\% CI should contain the true missing data with probability 95\%. 
In general, these CI are not symmetric around the imputed value
due to the nonlinear transformation $\bigf$.
We will continue to use the white wine dataset to illustration.
After fitting the Gaussian copula model, we can obtain the imputation CI as shown below.
\begin{CodeChunk}
\begin{CodeInput}
>>> ct = model_wine.get_imputed_confidence_interval()
>>> upper, lower = ct['upper'], ct['lower']
\end{CodeInput}
\end{CodeChunk}
By default, the method \code{get_imputed_confidence_interval()} extracts the imputation CI of the data used to fit the Gaussian copula model, with significance level \code{alpha=0.05}.
The empirical coverage of the returned CI is $0.943$, as shown below. 
Hence we see the constructed CI are well calibrated on this dataset.
\begin{CodeChunk}
\begin{CodeInput}
>>> missing = np.isnan(X_wine_masked)
>>> X_missing = X_wine[missing]
>>> cover = (lower[missing]<X_missing) & (upper[missing]>X_missing)
>>> np.round(cover.mean(),3)
\end{CodeInput}
\begin{CodeOutput}
0.943
\end{CodeOutput}
\end{CodeChunk}
The default setting uses an analytic expression to obtain the CI. 
As in \cref{sec:imputation}, when some variables are not continuous,
a safer approach builds CI using empirical quantiles computed from multiple imputed values.
Let us now construct the quantile CI and compare them with the analytical counterparts.
As shown below, the quantile CI has almost the same empirical coverage rate as the analytical CI, validating that the CI are well calibrated.
\begin{CodeChunk}
\begin{CodeInput}
>>> ct_q = model_wine.get_imputed_confidence_interval(type='quantile')
>>> upper_q, lower_q = ct_q['upper'], ct_q['lower']
>>> cover_q = (lower_q[missing]<X_missing) & (upper_q[missing]>X_missing)
>>> np.round(cover_q.mean(),3)
\end{CodeInput}
\begin{CodeOutput}
0.942
\end{CodeOutput}
\end{CodeChunk}
\section{Concluding remarks}
\gcpackage supports a variety of missing data imputation tasks including single imputation, multiple imputation, imputation confidence intervals, as well as imputation for large datasets and streaming datasets.
As a complement to this article, we  provide a \proglang{R} package\footnote{https://github.com/udellgroup/gcimputeR} and usage vignettes\footnote{https://github.com/udellgroup/gcimpute/blob/master/Examples}  detailing more specific topics such as trouble shooting, relative reliability for a single imputation, etc.

Although this article focuses on missing data imputation, 
\gcpackage can also be used to fit a Gaussian copula model to complete mixed datasets.
The resulting latent correlations may be useful to understand 
multi-view data collected on the same subjects from different sources.
As far as we know, no other software supports Gaussian copula estimation 
for mixed continuous, binary, ordinal and truncated variables.
\citet{fan2017high} only supports continuous and binary mixed data;
\citet{feng2019high} supports continuous, binary and ordinal mixed data; \citet{yoon2020sparse} supports continuous, binary and zero-inflated (a special case of truncated) mixed data.

One major area for future research is the appropriate treatment of nominal values.
\gcpackage currently encodes nominal variables using a one-hot encoding,
however, this encoding is not self-consistent for the copula model
since our estimation procedure ignores the one-hot constraint.
We advise users to model their features directly as ordinal or binary, if possible.

\gcpackage estimates the model provably well when data is missing uniformly at random (MCAR), and can estimate the copula provably well given the marginals 
if the data is missing at random (MAR).
Adapting the theory to handle data missing not at random (MNAR) is challenging.
However, we find empirically that \gcpackage still performs reasonably well in this setting.
Indeed, many different missing patterns may be called MNAR,
and imputation methods designed for one MNAR mechanism do not necessarily outperform on other MNAR data due to this heterogeneity.
We advise users to make the choice by evaluating on a validation dataset.

\section*{Acknowledgments}
The authors gratefully acknowledge support from NSF
Award IIS-1943131, the ONR Young Investigator Program,
and the Alfred P. Sloan Foundation. Special thanks to Xiaoyi
Zhu for her assistance in creating our Figure 1.

\bibliography{refs}
\end{document}